\documentclass[runningheads]{llncs}
\usepackage{graphicx}
\usepackage[ruled,vlined,linesnumbered,noresetcount]{algorithm2e}
\usepackage{float}
\begin{document}
\title{Parallel Algorithms for Densest Subgraph Discovery Using Shared Memory Model}
%
%
\author{B.D.M. De Zoysa\inst{1} \and
Y.A.M.M.A. Ali\inst{1} \and
M.D.I. Maduranga\inst{1} \and
Indika Perera\inst{1} \and
Saliya Ekanayake\inst{2}
Anil Vullikanti\inst{3}}
\authorrunning{B.D.M. De Zoysa et al.}
%
\institute{Department of Computer Science and Engineering, University of Moratuwa, Moratuwa, Sri Lanka\\ \email{\{dinukamanohara.15,ashkar.15,isurumaldeniya.15, indika\}@cse.mrt.ac.lk} \and
Lawrence Berkeley National Lab, California, USA\\
\email{\{esaliya\}@gmail.com} \and
University of Virginia, USA\\
\email{\{asv9v\}@virginia.edu}}
\maketitle              
\begin{abstract}
The problem of finding dense components of a graph is a widely explored area in data analysis, with diverse applications in fields and branches of study including community mining, spam detection, computer security and bioinformatics. This research project explores previously available algorithms in order to study them and identify potential modifications that could result in an improved version with considerable performance and efficiency leap. Furthermore, efforts were also steered towards devising a novel algorithm for the problem of densest subgraph discovery. This paper presents an improved implementation of a widely used densest subgraph discovery algorithm and a novel parallel algorithm which produces better results than a 2-approximation.

\end{abstract}
\section{Introduction}
\subsection{Background}
Discovering the dense components of a graph has become an important primitive in data analysis over the past few decades. Numerous algorithms and techniques had been developed and they have evolved throughout the period in order to ascertain the adequacy of methods to determine the dense components of different types of graphs namely, undirected, directed and weighted graphs with minimal requirement of resources. One such approach to discover dense components of a graph is to find the densest subgraph of a given graph.

By observing the applications that exploit the methods of finding densest subgraphs, one can easily acknowledge the substantial impact these methods have over modern technology due to their supplementary nature with well-known fields of research such as bioinformatics, data science and network analysis. Even though researchers have developed a variety of methods to perform the task of finding the densest subgraph of a graph, we could discern that there are areas yet to be explored and implementations that could be improved such that resource and time consumption can be mitigated further. This research focuses on improved methods for undirected graphs which can be modified for other types of graphs.

\subsection{Preliminaries}
\begin{definition}
Edge density \(\rho\textit{(G)}\) of a graph \(\textit{G(V,E)}\) is defined as \(\rho\textit{(G)}=\frac{|E|}{|V|}\). V is the set of vertices and E is the set of edges induced by V.
\end{definition}
\begin{definition}
The k-core of a graph is the largest subgraph in which every vertex has degree at least k.
\end{definition}
\begin{definition}
For \(\alpha \geq \textit{1}\), an algorithm is said to obtain an \(\alpha\textit{-approximation}\) to the undirected densest subgraph if it outputs a set \(\textit{\~{S}}\subseteq\textit{V}\) such that \(\rho\textit{(}\textit{\~{S}}\textit{)}\geq \frac{\rho^\textit{*}\textit{(G)}}{\alpha}\)~\cite{6}. \(\rho^\textit{*}\textit{(G)}\) is the density of the densest subgraph of graph \(\textit{G(V,E)}\).
\end{definition}

\subsection{Contributions}
Our main contributions are,
\begin{enumerate}
    \item Highly scalable shared memory implementation of densest subgraph approximation algorithm, Algorithm 1 in Bahmani et. al~\cite{6}. (In this paper we call this algorithm as Bahmani's algorithm.)
    \item A new highly scalable parallel algorithm which calculates more accurate density values than 2-approximation algorithms.
\end{enumerate}

\section{Related Work}
\vspace{-2.5mm}
Dense subgraph discovery in graphs has extensively been studied\cite{31,56,11,66}. The algorithms for finding the dense components in a graph can be categorized as exact enumeration, fast heuristics, and bounded approximation algorithms~\cite{x4}. 

The simplest and the most common way to define the density of an undirected graph is the edge density. The densest subgraph of a given graph based on the edge density is the subgraph that has the highest edge density than all the other possible subgraphs. Goldberg et al. presented a maximum-flow based exact algorithm to calculate that densest subgraph ~\cite{32}. Finding densest subgraphs with size restrictions (i.e., densest subgraphs which consist of a specified number of vertices) is a variant of this edge density densest subgraph problem and identified as NP-hard~\cite{5,4}. Another variant is the optimal quasi-clique problem~\cite{66}. This is also NP-hard~\cite{9}. Solutions for finding the top-k locally densest subgraphs were presented by Qin et al.~\cite{54}. Densest subgraph problem induced by evolving graphs is studied in~\cite{19}. Graph decomposition based on the edge density is explored in~\cite{64,18}. The density for directed graphs is expatiated in~\cite{43} and the densest subgraph problem on directed graphs is studied in~\cite{10}.

Approximation algorithms and algorithms based on heuristics to discover the densest subgraphs come into the picture when the exact algorithms fail to perform efficiently on large graphs. A greedy 2-approximation algorithm is presented in~\cite{10}. Bahmani et al.~\cite{6} presented (\(2+2\epsilon\))-approximation algorithm (\(\epsilon>0\)) and this is the algorithm that is being parallelized in this project. There exists an approximation algorithm to discover the densest subgraph in directed graphs~\cite{44}. The h-clique densest subgraph problem is defined and studied by Tsourakakis et al.~\cite{65,49}. Top-k local triangle-densest subgraph discovery is based on the 3-clique and has been studied in~\cite{57}. Fang et al.~\cite{fang} proposed exact and approximation algorithms for h-clique (\(\textit{h}\geq2\)) densest subgraph problem based on core decomposition. Two prime issues of previous algorithms namely, resulting in isolated graphs and neglecting of important features of natural graphs are addressed by Bo Wu and Haiying Shen~\cite{x11}. They present a heuristic algorithm for massive undirected graphs and an exact algorithm for big data in order to handle the aforementioned issues.

Rozenshtein et al.~\cite{x12} study the problem of dynamic dense subgraph whose edge occur in a short time interval or equivalently, are temporary compact. This problem is NP-hard~\cite{x12}. McGregor et al.~\cite{x13} discuss the problem of approximating the densest subgraph in the dynamic graph stream model. They also present the first algorithm for approximating the density of the densest subgraph up to (\(1+\epsilon\)) factor in the dynamic graph stream model.

K-core~\cite{7,47,51,23,22,20,25,26,67,12}, k-truss~\cite{15,37,69,39,38}, k-(r, s) nucleus~\cite{60,58,61,59}, k-clique~\cite{16,34}, k-edge connected components~\cite{35,36} and k-plexes~\cite{63} can also be identified as the other formations of dense graph component models. Tatti et al.~\cite{x35} proved that the largest core of a graph is a 2-approximation to the densest subgraph.

\section{Methodology}

\subsection{Parallelization of Bahmani’s Algorithm}
\subsubsection{Introduction} Bahmani et al. presents a greedy approximation algorithm for finding densest subgraph of an undirected graph. For any \(\epsilon>0\), their algorithm makes \(\textit{O}(\log_{1+\epsilon}\textit{n})\) passes over the input data and finds a subgraph of which the density is guaranteed to be within a factor of (\(2+2\epsilon\)) of the optimum. As the authors claim, the algorithm is inherently parallelizable with minimal data dependencies between parallelized components.

As provided in their description, the algorithm takes an undirected graph \(\textit{G(V,E)}\) as input. It proceeds in passes while removing a constant fraction of the remaining nodes in each pass. They have also proved that one of such intermediate subgraphs forms a (\(2+2\epsilon\))-approximation to the densest subgraph. Furthermore, they note that, via flows or linear programming densest subgraph problem in undirected graph can be solved in polynomial time. Conversely, they have observed that flow and linear programming techniques scale poorly to internet sized graphs. However, despite the worst-case examples, they have established the evidences for their claim of their algorithms having the capability of yielding near-optimal solutions and being simpler and highly efficient than flow or linear programming-based techniques.

\subsubsection{Approach} 
\begin{figure}
    \begin{minipage}{0.33\textwidth}
     \centering
     \includegraphics[width=\linewidth]{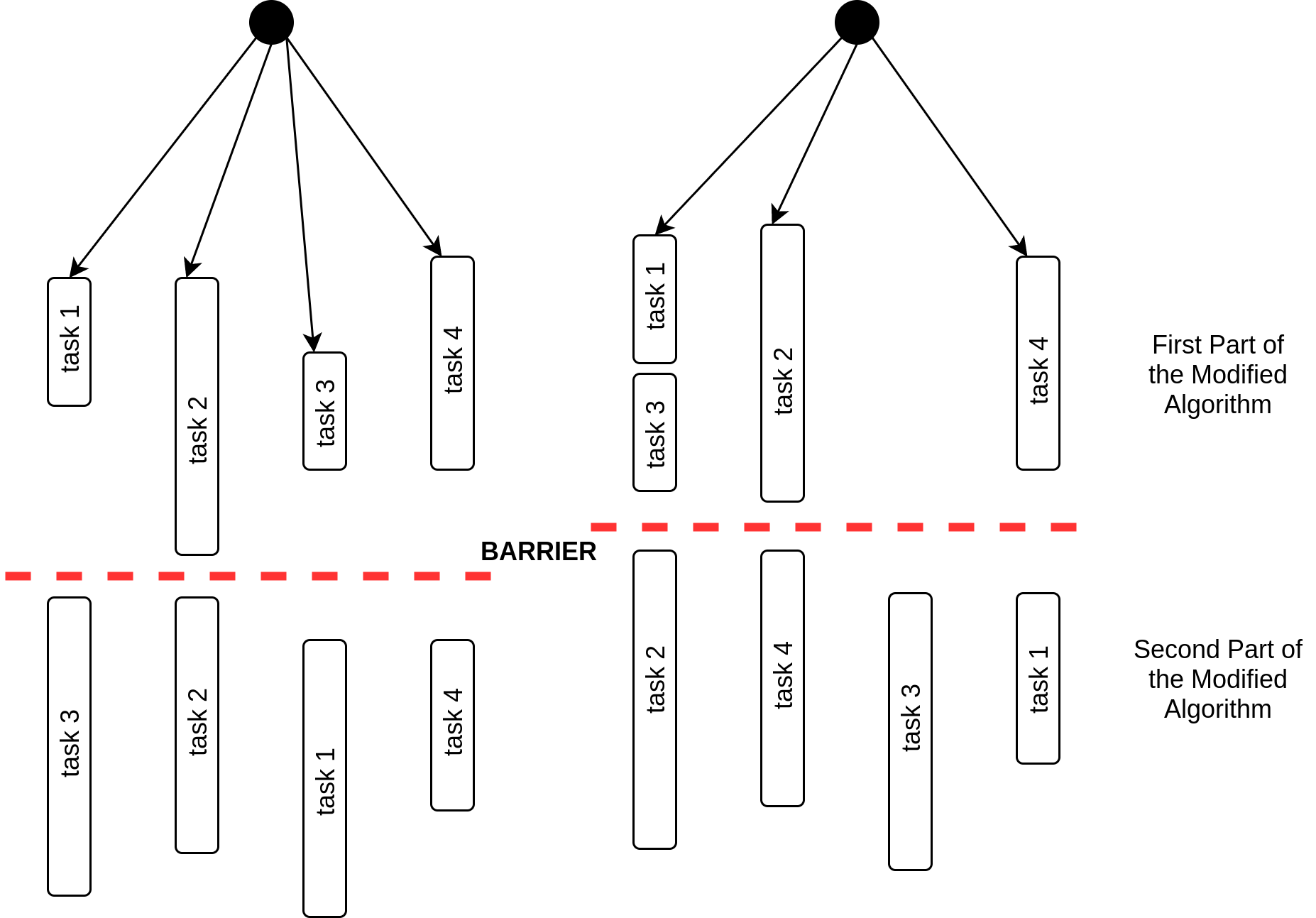}
     \caption{Tasks, parts and barrier} \label{fig1}
   \end{minipage}\hfill
   \begin{minipage}{0.33\textwidth}
     \centering
     \includegraphics[width=.7\linewidth]{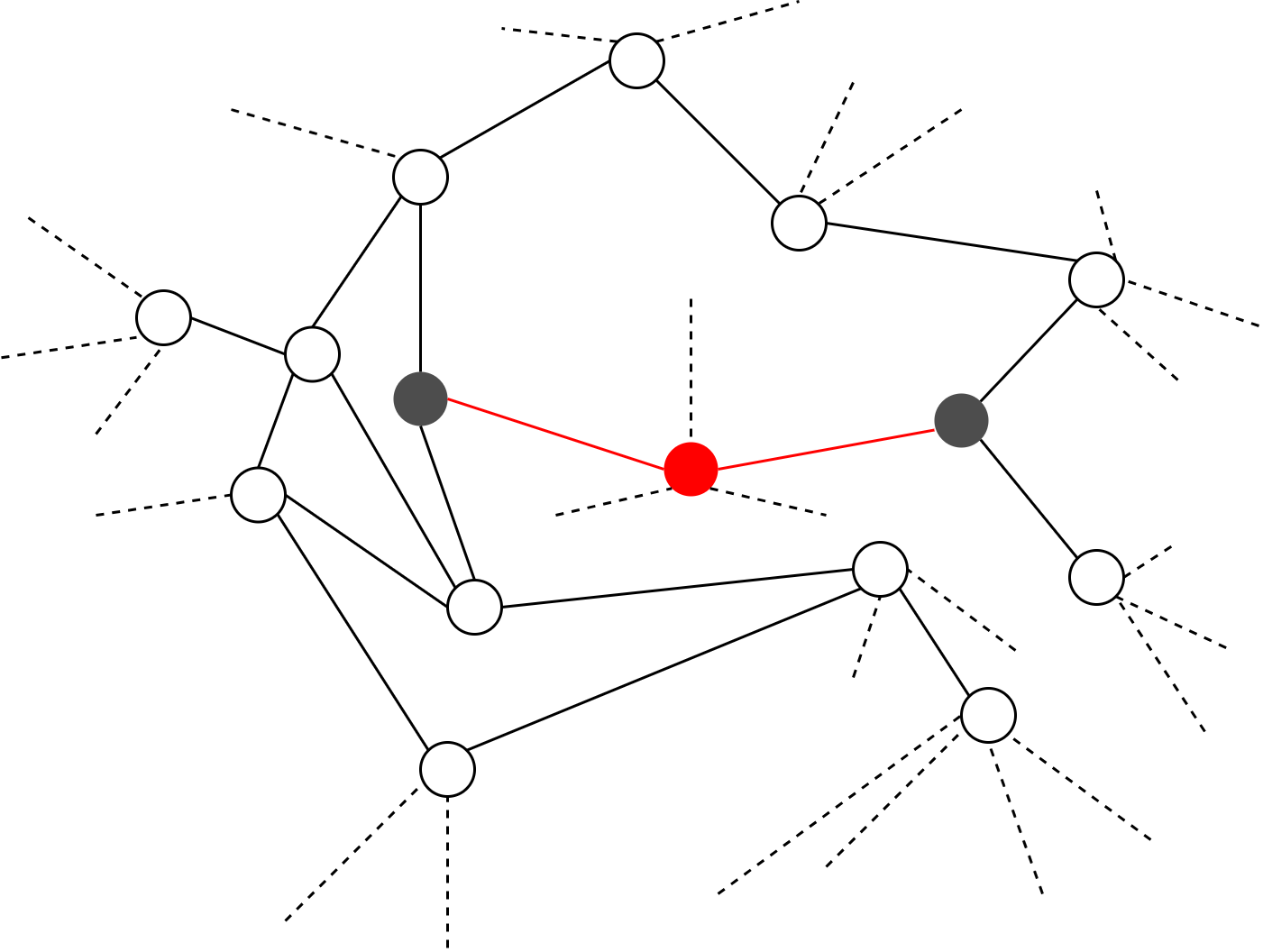}
     \caption{Critical node}\label{fig2}
   \end{minipage}
    \begin{minipage}{0.33\textwidth}
     \centering
     \includegraphics[width=\linewidth]{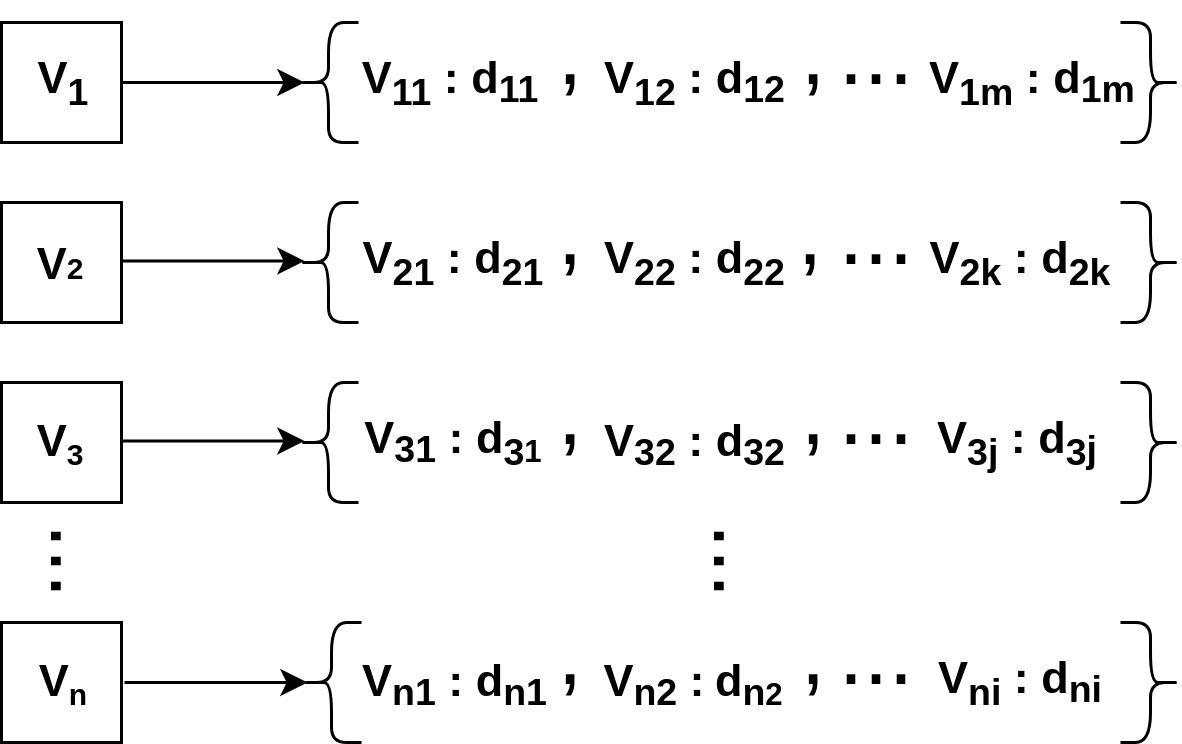}
     \caption{Graph data structure} \label{fig3}
   \end{minipage}\hfill
\end{figure}

The approach can be broken down mainly into two parts as described below:
\paragraph{Exploitation of task parallelism:} Tasks were introduced in version 3.0 of OpenMP. Before the availability of tasks, it was possible to execute loops of which the length was known at run time. In addition, there was also an imposed limit in the number of parallel sections. Most of the times this did not work well with even some well- known problems. i.e., even though it was possible to come with a feasible workaround, the solution was often unattractive. Until the introduction of tasks, it was not possible to efficiently and easily implement several types of parallelism. With tasks the programmer has to specify where the tasks are; then the OpenMP runtime system will generate a new task when a thread encounters a task construct. It is up to the OpenMP runtime system to ‘decide’ the moment of execution of the task whereas it can be delayed or immediate. task synchronization is used to enforce the completion of a task. Real world graphs are often highly unbalanced. Due to the nature of task creation, execution and its properties, it is plain that OpenMP tasks are highly applicable to the problem at hand. This is because if we use other patterns such as loop parallelism or work sharing constructs, the parallelized components that have comparatively lower amount of work to do have to be idle for most of the execution time. On the contrary, if the pattern used is task parallelism then the aforementioned occurrence would not be the case as tasks are not always unique to threads; consequently, a task can take over another component in the queue as soon as its work is done as opposed to being idle. In a considerable number of cases this could reduce the overhead of thread creation while helping execute the program quicker than other cases. The component above the dashed line representing barrier synchronization in figure~\ref{fig1} diagrammatically depicts this setting for a program with four tasks.

\paragraph{Partitioning the algorithm into two major parts:} First major part of the modified algorithm can execute without any synchronizations between the parallelized components. When it comes to the second major part, a synchronization is required due to the updating of degrees of vertices. Assuming that the threshold for removing vertices is between 3 and 4 in a given iteration, we can see in figure~\ref{fig2} that the grey shaded vertices should fail as their degree is lower than the threshold. If those vertices belong to two different tasks, then the neighboring red coloured critical vertex is vulnerable to race condition. This establishes the requirement for a synchronization.

\subsubsection{Implementation}
\paragraph{Data structures:} Despite relatively high memory usage we have used hash tables for storing the graph as our main focus was to be able to test the run-times for graphs of all types of vertex distribution. i.e., by doing this, we were able to support the graphs regardless of whether a graph of n vertices contains vertices from \(0\) to \((\textit{n}-1)\) or not. For this purpose, we have used the unordered map which is the C++ implementation of hash tables. We define a structure named super map which concretely represents the whole graph. Each key of the hash table represents a vertex of the graph. The value for each key is another hash table unordered map which has the neighboring vertices and their degrees as keys and values respectively.

\paragraph{Parallelization:} 

\begin{itemize}
    \item Task ID creation: As mentioned previously, tasks and threads does not always have one to one relationship. Therefore, as our program requires an ID unique to each task, the ID creation process is done by defining and initializing a shared variable ID and then atomically assigning the variable to the thread private ID immediately after each individual task commences. This is followed by increment of the shared variable which can then be assigned to the next task. The id creation process is identical in both the major parts of P-Bahmani.
    \item Parallelism in the first part of P-Bahmani: This part does not contain any inter-task data dependencies. Moreover, for the purposes of time reduction in execution of the program, we have exercised a strategy where the algorithm avoids the testing of already visited vertices by removing those vertices. Here, each parallelized component iterates over the set of locally assigned vertices and removes the failed vertices from the locally assigned active set of vertices. With this strategy, the algorithm does not need to go over the failed vertices in the upcoming iteration. This cannot be done with work sharing constructs or basic type of loop parallelism where the Internal Control Variables are bound to a thread because an increment clause of OpenMP for loop must only perform simple addition/subtraction on the loop variable. However, the kind of increment our program consists in order to perform the removal of elements and the iteration over the same set is case dependent. The for loop is defined only with the pointer initiation and the condition. But the method of the pointer assignment must be defined on the fly depending on whether the current vertex failed or not. This can only be realized with tasks.
    \item Parallelism in the second part of P-Bahmani: Each of the parallelized unit of executions contains a unique set of failed vertices resulted from the first part of the algorithm. When removing edges, we have to decrease the degrees of sources as well as the destinations. Therefore, while iterating over the unique set of task-local failed vertices, it is necessary to iterate over the set of neighbors of the current vertex since we have to decrease the degree of source vertex via the direction source\(\rightarrow\)destination and the degree of destination vertex via the direction destination\(\rightarrow\)source. Since the source vertex is unique to each task and both the aforementioned directions consists the source, we can vouch that the assignment of variables at the addresses of sources won’t encounter any race conditions. Hence, the task local edge reduction variable can safely be used for decreasing the number of edges. Conversely, it is obvious that the destination vertex is vulnerable to race conditions which dictates us to use a critical section when updating its degree. We have tackled this through the C++’s built-in function for atomic reductions. The second part of the algorithm has the form shown in figure~\ref{fig4}.
\end{itemize}

\RestyleAlgo{ruled}
\begin{algorithm}
\caption{P-Bahmani}
\SetAlgoLined
\LinesNumbered
\SetKwInOut{Input}{Input}
\Input{\(\textit{G(V,E)}, \epsilon > 0\)}
\(\tilde{G}\leftarrow G\)\;
\(n_{e}, n_{v} \leftarrow 0;\)$\hfill
\triangleright Task-local\hspace{0.1cm}variables\hspace{0.1cm}of\hspace{0.1cm}type\hspace{0.1cm}\textit{int}$
 
\(active; failed;\)$\hfill\mathrm{\triangleright Task-local\hspace{0.1cm}variables\hspace{0.1cm}of\hspace{0.1cm}type\hspace{0.1cm}\textit{set}}$
 
\(\rho,\tilde{\rho} \leftarrow \rho_0;\)$\hfill\mathrm{\triangleright Global\hspace{0.1cm}variables\hspace{0.1cm}of\hspace{0.1cm}type\hspace{0.1cm}\textit{set}}$

\While{\(G \rightarrow |V| > 0\)}{
    \(\textbf{parallel}\)
    \For{\(\nu \in active\)}{
        \uIf{\(\nu.deg \leq 2(1+\epsilon)\rho\)}{
            \(failed.insert(\nu)\)\;
            \(n_{v} \leftarrow n_{v} - 1\)\;
            \(active.remove(\nu)\)\;
        }\Else{
            \(pass\)\;
        }
    }
  
    \(\textbf{barrier synchronization}\)
    
    \(\textbf{parallel}\)
    \For{\(\nu \in failed\)}{
        \For{\(u \in \nu.neighbours\)}{
            \(n_{e} \leftarrow n_{e} - u.local\_deg\)\;
            \(\textbf{atomicSub}(u.deg, u.local\_deg)\)\;
        }
    }
  
    \(\textbf{reduce}(G \rightarrow |V|, n_{v}); \textbf{reduce}(G \rightarrow |E|, n_{e})\)\;
    \(n_{v},n_{e} \leftarrow 0\)\;
    \(failed.clear()\)\;
  
    \uIf{\(\rho > \tilde{\rho}\)}{
        \(\tilde{G} \leftarrow G\)\;
    }\Else{
        \(pass\)\;
    }
}
\(\textbf{Return: }\tilde{G}\)\;
\end{algorithm}


\vspace{5mm}
\subsection{Core-based Dense Subgraphs}
\subsubsection{Approach} We previously described how the densest subgraph is correlated to the densest k-core (\({\textit{S}_\textit{k}}^*\)). Identification of the densest k-core provides us with a 2-approximation to the densest subgraph. This connotes that if we attempt to improve the density of the densest k-core, we will end up with a better approximation to the densest subgraph.

\begin{figure}
   \begin{minipage}{0.33\textwidth}
     \centering
     \includegraphics[width=\linewidth]{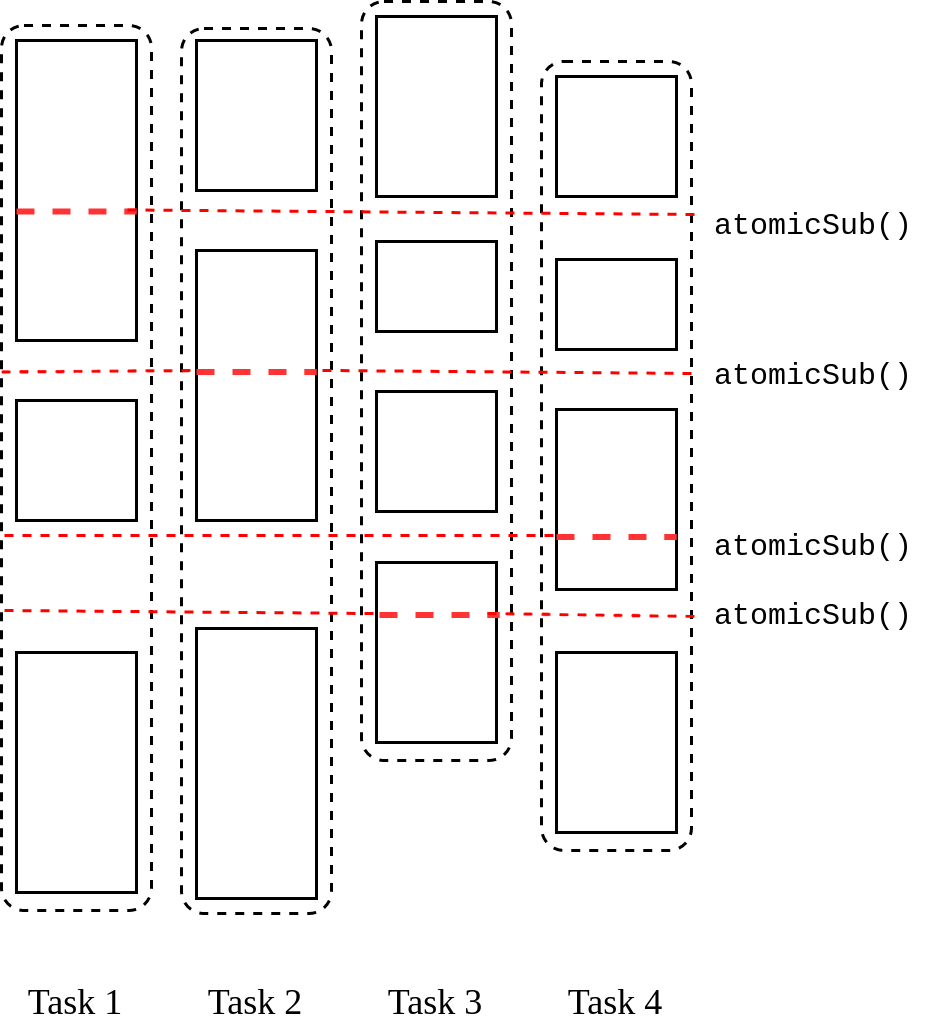}
     \caption{Second part of P-Bahmani with synchronization} \label{fig4}
   \end{minipage}
    \begin{minipage}{0.33\textwidth}
     \centering
     \includegraphics[width=\linewidth]{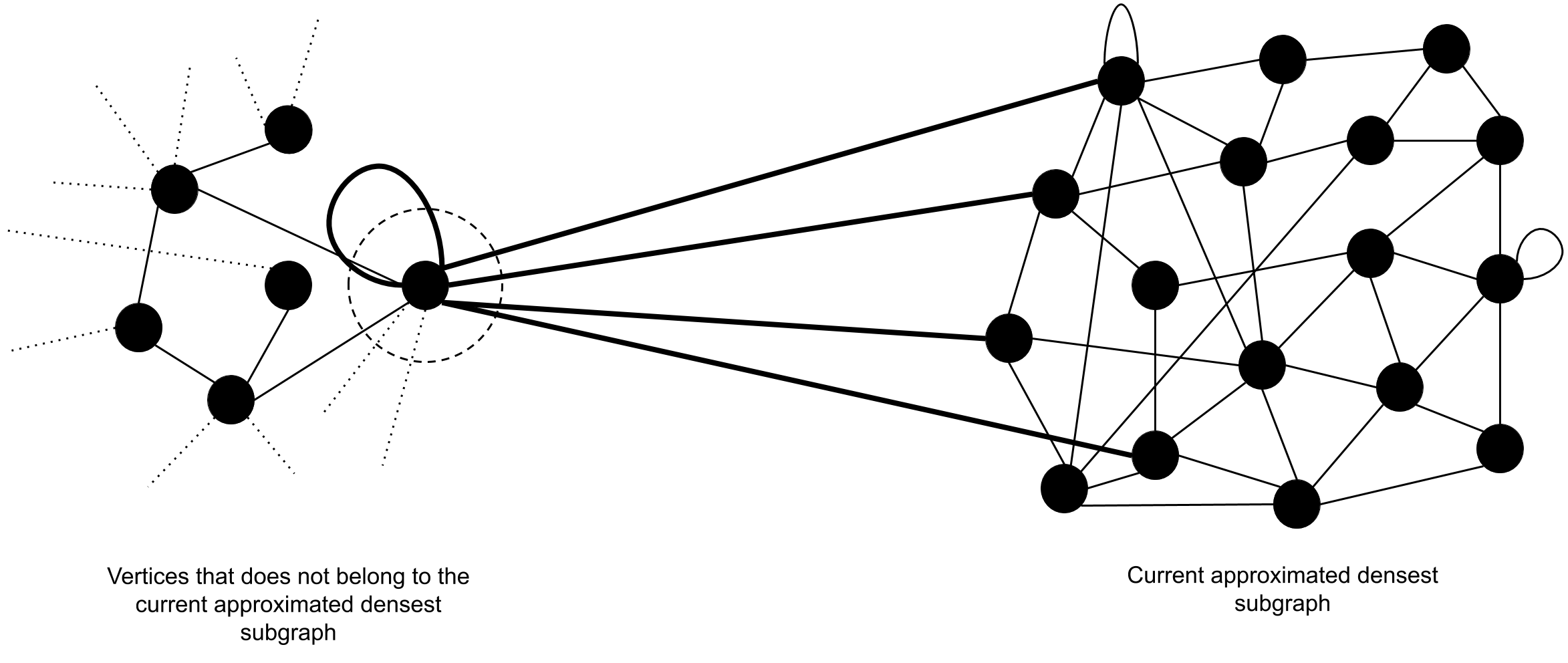}
     \caption{} \label{fig5}
   \end{minipage}\hfill
   \begin{minipage}{0.33\textwidth}
     \centering
     \includegraphics[width=\linewidth]{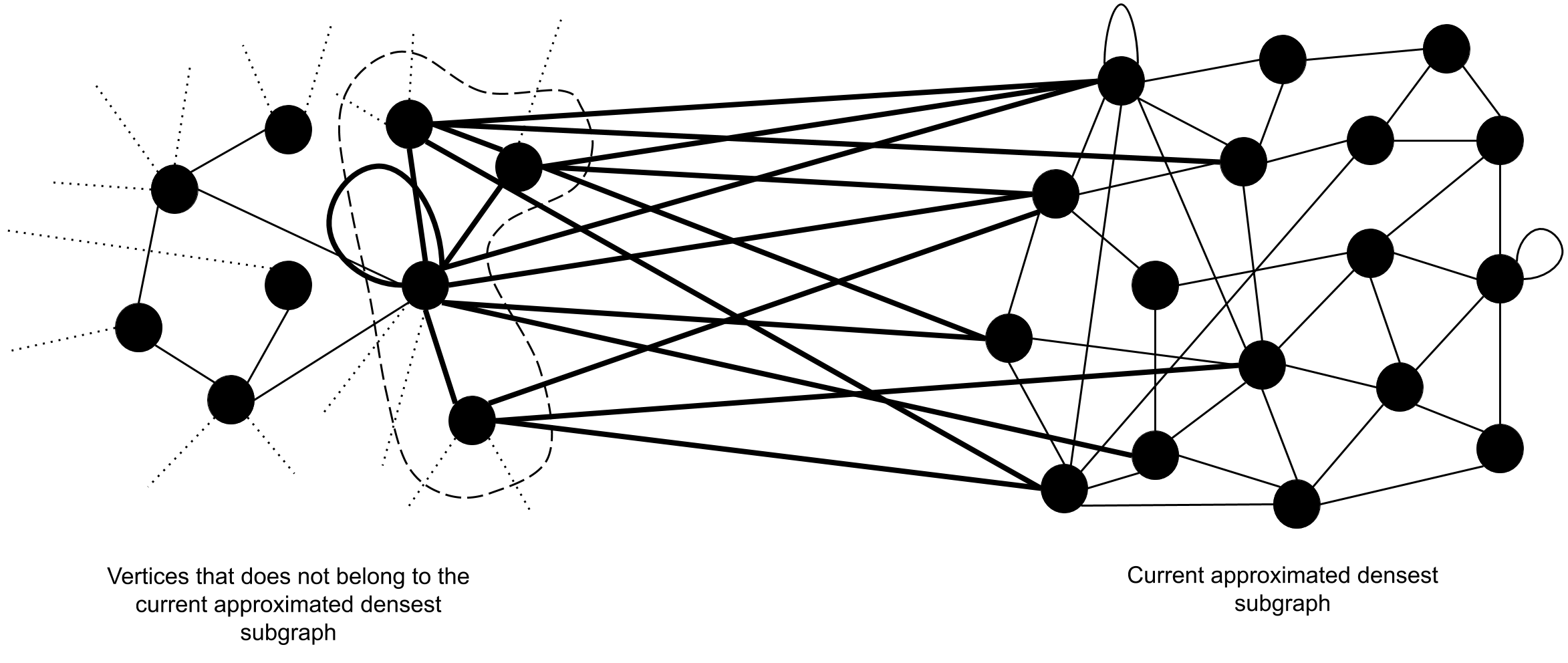}
     \caption{} \label{fig6}
   \end{minipage}
\end{figure}

The notion of improving the density of \({\textit{S}_\textit{k}}^*\) can be realized by adding new vertices from the set of vertices that does not belong to \({\textit{S}_\textit{k}}^*\). We show that the density of a subgraph \(\tilde S\) can be increased by adding appropriately chosen vertices where \(\tilde S\) is the subgraph which accommodates the newly added vertices, starting from \({\textit{S}_\textit{k}}^*\). Therefore, we introduce a method of selecting outside vertices which would increase the density of \(\tilde S\) when appended. This method is based on the density of \(\tilde S\), \(\rho(\tilde S)\) and the number of edges \(\tilde e\), induced by the set of vertices inside \(\tilde S\) and the vertex in consideration of. As depicted by figure~\ref{fig5} and figure~\ref{fig6}, the number edges marked in thick lines should be counted as \(\tilde e\). It is obvious that if we add a new vertex that introduces \(\tilde e\) number of suitable edges such that \(\tilde e>\rho(\tilde S)\), then \(\rho(\tilde S)\) is increased by an amount of \((\textit{n} \tilde e-\textit{e})/(\textit{n}(\textit{n}+1))(>0)\), where \(\textit{n}\) is the number of vertices and \(\textit{e}\) is the number of edges inside \(\tilde S\) before the addition of the new vertex.

Our approach towards designing the algorithm consists of two distinct procedures. They are,
\begin{enumerate}
    \item Finding the densest k-core using an appropriate k-core decomposition algorithm. Here we have to use two different decomposition algorithms for serial and parallel versions.
    \item Selecting suitable vertices which can be added to \(\tilde S\) and appending them to \(\tilde S\).
\end{enumerate}
\vspace{-5mm}

\subsubsection{CBDS-P Algorithm} When designing the algorithm, we utilize the parallel k-core decomposition algorithm, PKC, proposed by Kabir et al.~\cite{x34}.

CBDS-P takes a graph and outputs an approximate densest subgraph. The first phase of the algorithm which is based on PKC algorithm, determines the core-ness values of the vertices. Apart from calculating the core-ness values, the first phase is modified to cater the self-edges of vertices and to calculate the densities of detected cores.

The second phase of CBDS-P detects suitable vertices outside the densest core. Threads find vertices which has a higher degree than the density of the densest core and a lesser degree than the core-ness value of the densest core. Then these selected vertices are processed in order to find vertices which possess legitimate edge counts that satisfy the conditions presented in the beginning of this subsection. We call these vertices, legitimate vertices. Ultimately the legitimate vertices’ degrees are set to \(\textit{max\_density\_core}\) and the maximum density is calculated. Algorithm 2 returns the maximum density (\(\textit{max\_density}\)), maximum density core value (\(\textit{max\_density\_core}\)) and the input graph. All the vertices that belong to the approximate densest subgraph can be identified by checking their degree values, degree values are greater than or equals to \(\textit{max\_density\_core}\). Note that some vertices may have a degree value equal to the \(\textit{max\_density\_core}\) even though they do not belong to the densest core (i.e., here \(\textit{max\_density\_core}\) value also serves as a labelling value).

First phase of CBDS-P has the time complexity equivalent to the time complexity of PKC. It is \(O(|V|K_{max}+|E|)\). Here, \(K_{max}\) is the largest core value. In second phase, we go through the complete graph in order to detect the suitable vertices, adding a \(O(|V|)\) complexity. It takes another \(O(|V^{'}|)\) to go through \(|V^{'}|\) number of neighbours introduced by the detected eligible vertices. Finally, to calculate the number of edges between the legitimate vertices, it takes \(O(|V^{''}|^{2}-|V^{''}|)/2\). \(|V^{''}|\) is the number of legitimate vertices. Therefore, the time complexity of CBDS-P is \(O(|V|(K_{max}+1)+|V^{'}|+|V^{''}|^{2}+|E|)\). Based on our experimental results, it is evident that \(|V^{'}|\ll|V|\) and \(|V^{''}|\ll|V|\).

To implement CBDS-P, we followed the same design concepts and technologies we used to implement P-Bahmani. 

\RestyleAlgo{ruled}
\begin{algorithm}
\caption{CBDS-P}
\SetAlgoLined
\LinesNumbered
\SetKwBlock{Begin}{Begin}{}
\SetKwInOut{Input}{Input}
\Input{\textit{G(V, E)}}
\Begin{
\(visited\leftarrow0; max\_density\leftarrow0; max\_density\leftarrow0; max\_density\_core\leftarrow0\)\;
\(intermediate\_edges\leftarrow0; m_{v}\leftarrow0; m_{e}\leftarrow0\)\;
\(legit\_vector;\)$\hfill
 \mathrm{\triangleright Global\hspace{0.1cm}vector\hspace{0.1cm}of\hspace{0.1cm}type\hspace{0.1cm}\textit{int}}$

\(\textbf{parallel do}\)

\(k\leftarrow0; start\leftarrow0; end\leftarrow0; deleted\_local\leftarrow0; aux\_local\leftarrow0; density\leftarrow0\)\;
\(buff;\)$\hfill
 \mathrm{\triangleright Thread-local\hspace{0.1cm}array\hspace{0.1cm}of\hspace{0.1cm}type\hspace{0.1cm}\textit{int}\hspace{0.1cm}with\hspace{0.1cm}size\hspace{0.1cm}|V|/N,\hspace{0.1cm}N=num\hspace{0.1cm}of\hspace{0.1cm}threads}$

\(eligible\_vector;\)$\hfill
 \mathrm{\triangleright Thread-local\hspace{0.1cm}vector\hspace{0.1cm}of\hspace{0.1cm}type\hspace{0.1cm}\textit{int}}$

\While{\(visited<G\rightarrow|V|\)}{
	\(\textbf{parallel}\)
    	\For{\(\nu\in G\rightarrow V\)}{
            \If{\(\nu.deg=k\)}{
                \(buff[end]\leftarrow\nu\)\;
                \(end++\)\;
		    }
        }
    \While{\(start<end\)}{
    	\(v\leftarrow buff[start]; deleted\_local\leftarrow deleted\_local+k; start++\);
    	
        \For{\(u\in\nu.neighbours\)}{
            \If{\(\nu\neq u\)}{
            	\If{\(u.deg>k\)}{
                	\(a\leftarrow\textbf{atomicSub}(u.deg, 1)\)\;
                    \If{\(a=k+1\)}{
                        \(buff[end]\leftarrow u; end++\)\;
                    }
                    \If{\(a\leq k\)}{
                    	\(\textbf{atomicAdd}(u.deg, 1)\)\;
                    }
                    \uIf{\(a>k\)}{
                    	\(aux\_local++\)\;
                    }\Else{
                        \(deleted\_local++; aux\_local--\);
                    }
                }
            }
        }
    }
    \(\textbf{atomicAdd}(visited, end); \textbf{atomicAdd}(deleted, deleted\_local)\)\; \(\textbf{atomicAdd}(aux, aux\_local)\)\;
    \(\textbf{barrier synchronization}\)
    
    \(\textbf{single do}\)
    
    \(density\leftarrow(G\rightarrow|E|-(deleted+aux)/2)/(G\rightarrow|V|-visited)\)\;
    \If{\(density>max\_density\)}{
    	\(max\_density\leftarrow density; m_{e}\leftarrow G\rightarrow|E|-((deleted+aux)/2)\)\; \(m_{v}\leftarrow G\rightarrow|V|-visited; max\_density\_core\leftarrow k\)\;
    }
    \(\textbf{end single}\)
    
    \(start\leftarrow0; end\leftarrow0; deleted\_local\leftarrow0; aux\_local\leftarrow0; k++\)\;
}
}
\end{algorithm}

\RestyleAlgo{ruled}
\SetNlSty{texttt}{(}{)}
\begin{algorithm}
\LinesNumbered
\setcounter{AlgoLine}{48}
\SetAlgoLined
\SetKwBlock{Begin}{}{End}
\Begin{
\(legits\leftarrow0\)\;
\(\textbf{parallel}\)
\For{\(\nu\in G\rightarrow V\)}{
	\If{\(\nu.deg>max\_density \textbf{ and } \nu.deg<max\_density\_core\)}{
		\(eligible\_vector.add(\nu)\)\;
	}
}
\(start\leftarrow0\)\;
\While{\(start<eligible\_vector.currentSize\)}{
	\(legits\leftarrow0; \nu\leftarrow eligible\_vector[start]\)\;
	\For{\(u\in\nu.neighbours\)}{
		\uIf{\(u.deg\geq max\_density\_core\)}{
			\(legits\leftarrow legits+1\)
		}\ElseIf{\(\nu=u\)}{
			\(legits\leftarrow legits+0.5\)
		}
	}
	\If{\(legits>max\_density\)}{
		\(\textbf{critical do}\)
		
		\(intermediate\_edges\leftarrow intermediate\_edges+legits\)\;
		\(legit\_vector.add(\nu)\)\;
		
		\(\textbf{end critical}\)
	}
	\(start++\)\;
}
\(\textbf{barrier synchronization}\)

\(i\leftarrow0\)\;
\(\textbf{parallel dynamic reduction(+: intermediate\_edges)}\)

\For{\(i<legit\_vector.currentSize-1\)}{
	\(\nu\leftarrow legit\_vector[i]; \nu.deg\leftarrow max\_density\_core; j\leftarrow i+1\)\;
	\For{\(j<legit\_vector.currentSize\)}{
		\(u\leftarrow legit\_vector[j]\)\;
		\If{\(u\in\nu.neighbours\textbf{ and }\nu\neq u\)}{
			\(intermediate\_edges\leftarrow intermediate\_edges+1\)\;
		}
		\(j++\)\;
	}
	\(i++\)\;
}
\(\textbf{barrier synchronization}\)

\(\textbf{single do}\)

\(m_{e}\leftarrow m_{e}+intermediate\_edges; m_{v}\leftarrow m_{v}+legit\_vector.currentSize\)\; \(max\_density\leftarrow m_{e}/m_{v}\)\;
\(\textbf{end single}\)

\(\textbf{end parallel}\)
}
\(\textbf{Return: } G, max\_density\_core, max\_density\)
\end{algorithm}

\section{Experimental Results and Evaluation}
\subsection{Datasets}
All the datasets that we have used to evaluate the performance of our algorithms are real-world datasets (natural graphs). They were obtained from the Stanford Network Analysis Project (SNAP) and are publicly available for anyone to use[]. Table~\ref{tab1} provides the details of the datasets. In order to ensure the accuracy of our results obtained from our implementations, we used datasets and results given in~\cite{6} as the reference. The table~\ref{tab2} provides the table from the paper. Each of the algorithms are evaluated based on accuracy and runtime.

\begin{table}
\caption{Dataset details}\label{tab1}
\centering
\begin{tabular}{|c|c|c|c|}
\hline
Dataset & Label & \(|V|\) & \(|E|\)\\
 \hline
 ca-GrQc  & GrQc    & 5,242 & 28,980\\
 ca-HepTh &   HepTh  & 9,877  & 51,971\\
 facebook-combined & Facebook & 4,039 &  88,234\\
 ca-HepPh   & HepPh & 12,008 &  237,010\\
 musae-squirrel-edges &  Squirrel  & 5,201 & 216,933\\
 as-skitter & Skitter  & 1,696,415  & 11,095,298\\
 soc-LiveJournal1 & LJ  & 4,847,571   & 68,993,773\\
 com-Orkut & Orkut & 3,072,441   & 117,185,083\\
\hline
\end{tabular}
\end{table}

\begin{table}[htp]
\caption{Reference data}\label{tab2}
\centering
\begin{tabular}{|c|ccc|c|ccc|}
\hline
\(\textit{G(V, E)}\) & \(|\textit{V}|\) && \(|\textit{E}|\) & \(\rho^*\textit{(G)}\) && \(\rho^*\textit{(G)}/\tilde{\rho}\textit{(G)}\) \cr &&&&& \(\epsilon = 0.005\) & \(\epsilon = 0.05\) & \(\epsilon = 0.5\)\\
\hline
 \textsc{as2000102} & 6,474 && 13,233 & 9.29 & 1.229 & 1.268 & 1.194\\
 \textsc{ca-AstroPh} & 18,772 && 396,160 & 32.12 & 1.147 & 1.156 & 1.273\\
 \textsc{ca-CondMat} & 23,133 && 186,936 & 13.47 & 1.072 & 1.072 & 1.429\\
 \textsc{ca-GrQc} & 5,242 && 28,980 & 22.39 & 1.000 & 1.000 & 1.395\\
 \textsc{ca-HepPh} & 12,008 && 237,010 & 119.004 & 1.000 & 1.017 & 1.151\\
 \textsc{ca-HepTh} & 9,877 && 51,971 & 15.50 & 1.000 & 1.000 & 1.356\\
 \textsc{email-Enron} & 36,692 && 367,662 & 37.34 & 1.058 & 1.072 & 1.063\\
\hline
\end{tabular}
\end{table}

\subsection{Testing Environment}
We used a m5.16xlarge Amazon EC2 which possess a 64 core Intel(R) Xeon(R) Platinum 8175M CPU @ 2.50GHz and a 246GB RAM.

\subsection{Density and Runtime Comparisons}
Table~\ref{tab3} contains the approximate maximum density values calculated by both P-Bahmani with \(\epsilon=0\) and CBDS-P. Note that when \(\epsilon=0\), P-Bahmani algorithm produces the highest accuracy and it is equivalent to 2-approximation algorithm by Charikar et. al[]. Therefore, in the table~\ref{tab3}, we can clearly observe that CBDS-P calculates more accurate maximum density values than a 2-approximation.

\begin{figure}
    \begin{minipage}{0.33\textwidth}
     \centering
     \includegraphics[width=\linewidth]{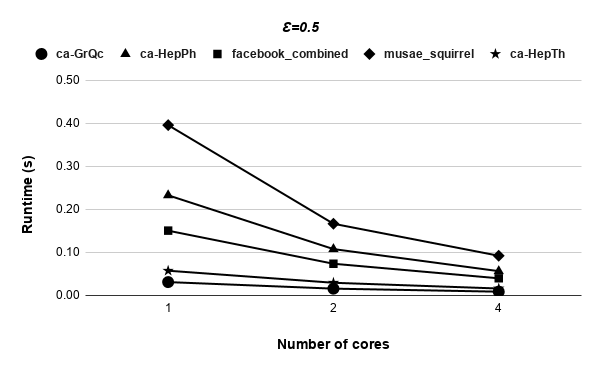}
     \caption{} \label{fig7}
   \end{minipage}\hfill
   \begin{minipage}{0.33\textwidth}
     \centering
     \includegraphics[width=\linewidth]{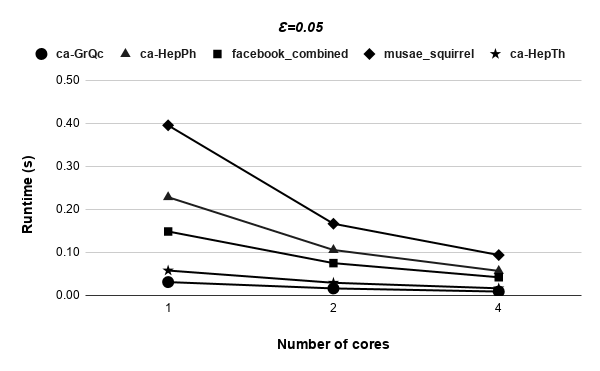}
     \caption{} \label{fig8}
   \end{minipage}\hfill
   \begin{minipage}{0.33\textwidth}
     \centering
     \includegraphics[width=\linewidth]{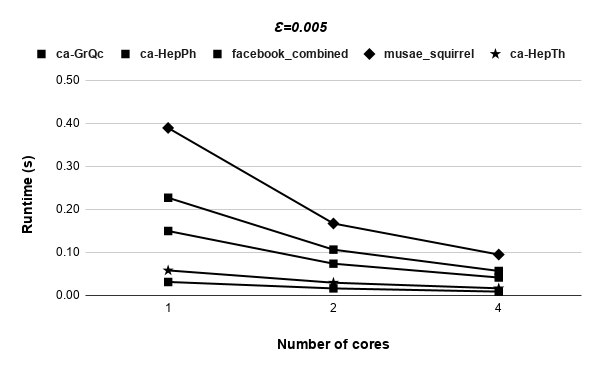}
     \caption{} \label{fig9}
   \end{minipage}
\end{figure}

\begin{figure}
    \begin{minipage}{0.33\textwidth}
     \centering
     \includegraphics[width=\linewidth]{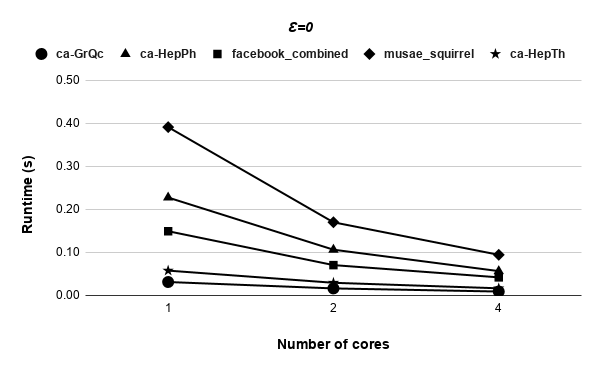}
     \caption{} \label{fig10}
   \end{minipage}\hfill
   \begin{minipage}{0.33\textwidth}
     \centering
     \includegraphics[width=\linewidth]{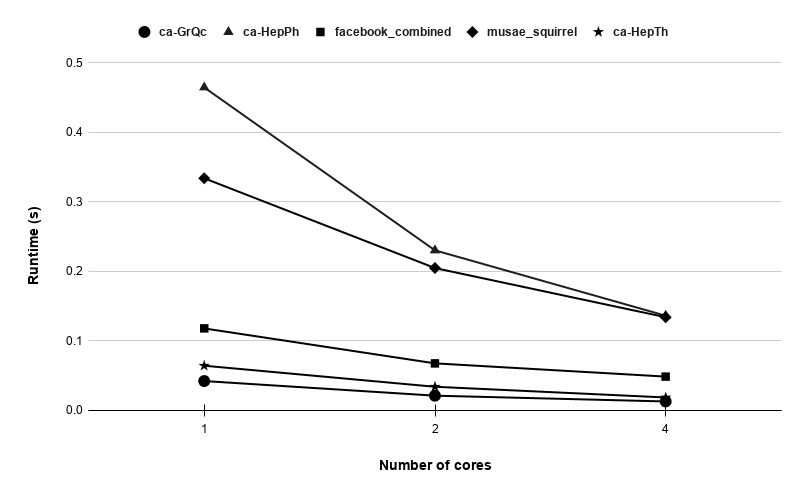}
     \caption{} \label{fig11}
   \end{minipage}\hfill
   \begin{minipage}{0.33\textwidth}
     \centering
     \includegraphics[width=\linewidth]{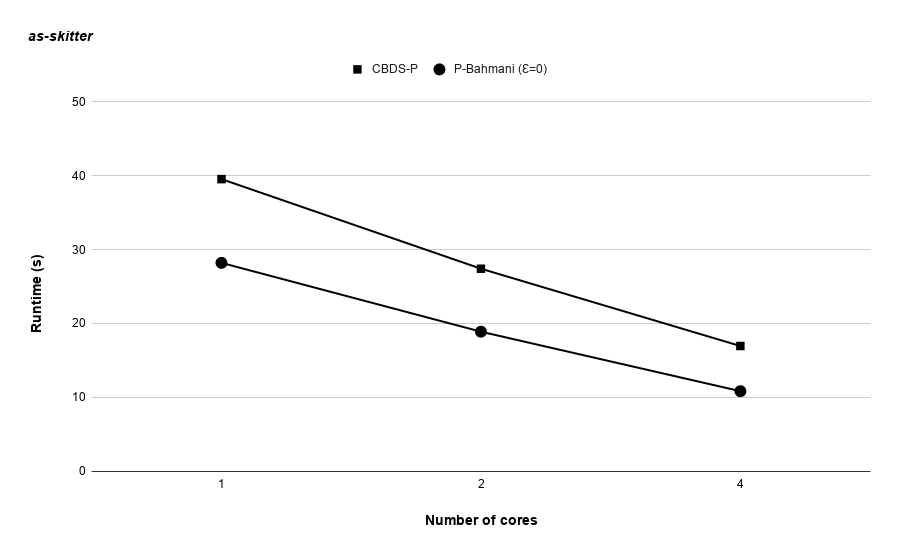}
     \caption{} \label{fig12}
   \end{minipage}
\end{figure}

\begin{figure}
    \begin{minipage}{0.33\textwidth}
     \centering
     \includegraphics[width=\linewidth]{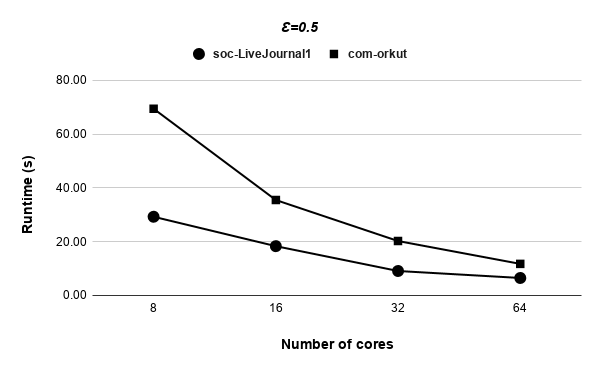}
     \caption{} \label{fig13}
   \end{minipage}\hfill
   \begin{minipage}{0.33\textwidth}
     \centering
     \includegraphics[width=\linewidth]{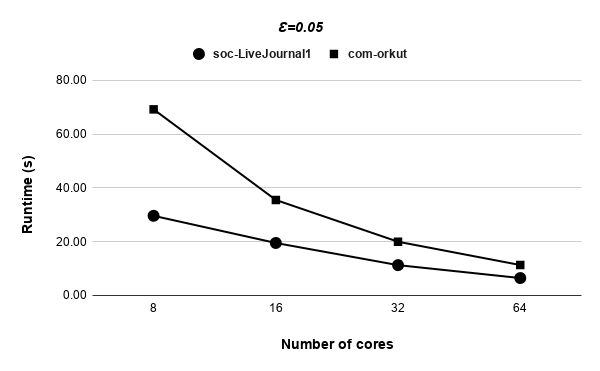}
     \caption{} \label{fig14}
   \end{minipage}\hfill
   \begin{minipage}{0.33\textwidth}
     \centering
     \includegraphics[width=\linewidth]{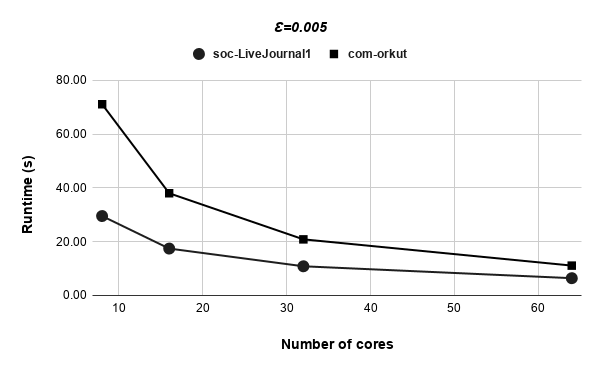}
     \caption{} \label{fig15}
   \end{minipage}
\end{figure}

\begin{figure}
    \begin{minipage}{0.48\textwidth}
     \centering
     \includegraphics[width=\linewidth]{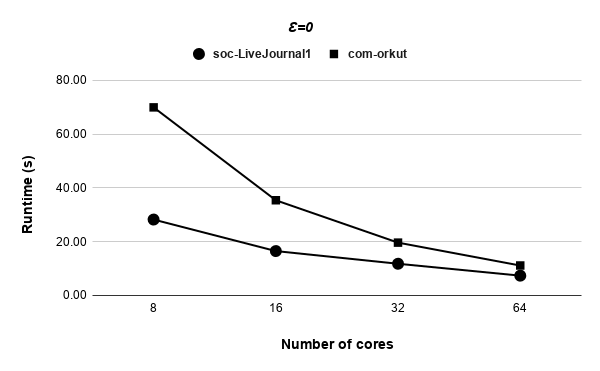}
     \caption{} \label{fig16}
   \end{minipage}\hfill
   \begin{minipage}{0.48\textwidth}
     \centering
     \includegraphics[width=\linewidth]{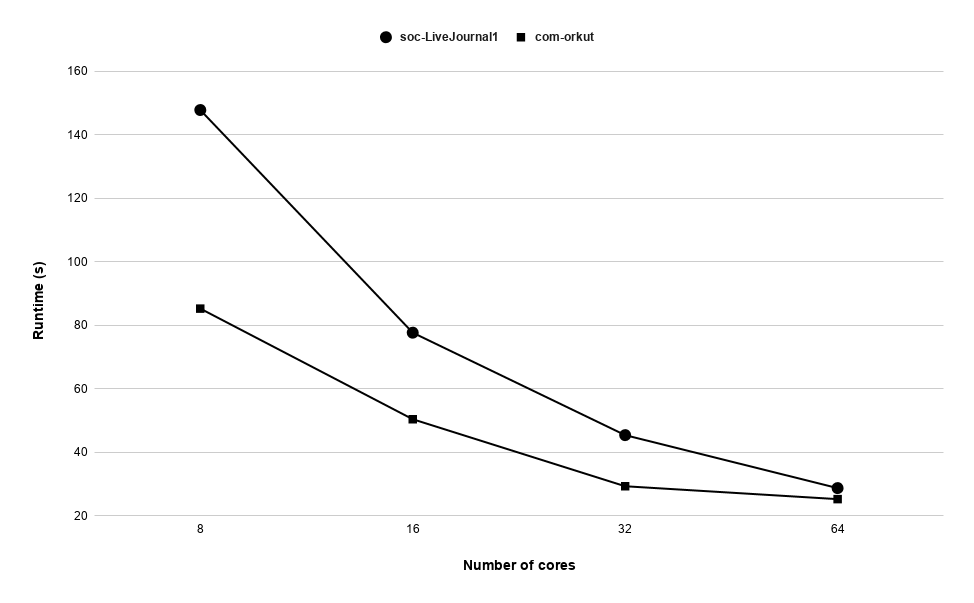}
     \caption{} \label{fig17}
   \end{minipage}
\end{figure}

Figures~\ref{fig7}, ~\ref{fig8}, ~\ref{fig9}, ~\ref{fig10} for first 5 graphs and figures~\ref{fig13}, ~\ref{fig14}, ~\ref{fig15}, ~\ref{fig16} for last 2 graphs are the runtime variation of P-Bahmani against number of cores for \(\epsilon\) values of 0.5, 0.05, 0.005 and 0 respectively. 

\begin{figure}
    \begin{minipage}{0.48\textwidth}
     \centering
     \includegraphics[width=\linewidth]{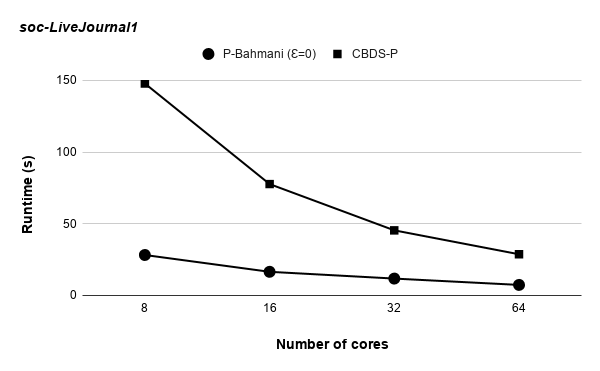}
     \caption{} \label{fig18}
   \end{minipage}\hfill
   \begin{minipage}{0.48\textwidth}
     \centering
     \includegraphics[width=\linewidth]{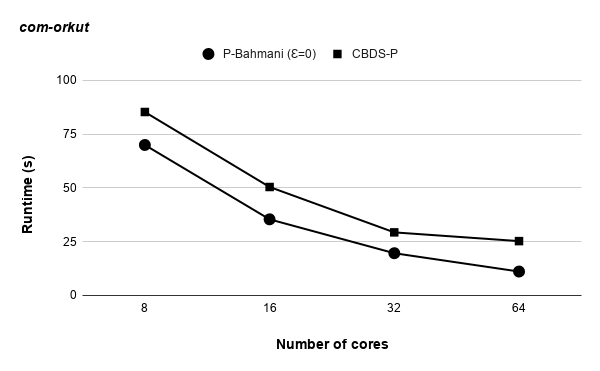}
     \caption{} \label{fig19}
   \end{minipage}
\end{figure}

Runtime variation of graphs as-skitter, soc-LiveJournal1 and com-orkut against the number of cores with respect to both P-Bahmani(\(\epsilon=0\)) and CBDS-P are shown in figures~\ref{fig12}, ~\ref{fig18} and ~\ref{fig19}. Further, figure~\ref{fig17} depicts the runtime of CBDS-P against the number of cores with respect to both soc-LiveJournal1 and com-orkut graphs.

\begin{table}[htp]
\caption{Density comparison}\label{tab3}
\centering
\begin{tabular}{|c|c|c|c|}
\hline
Dataset & Exact Density & P-Bahmani \((\epsilon=0)\) & CBDS-P \\
 \hline
 ca-GrQc  & 22.391    & 22.391 & 22.391\\
 ca-HepTh &   15.5  & 15.5  & 15.5\\
 facebook-combined & 77.347 & 69.9679 & 77.347\\
 ca-HepPh   & 119.004 & 119.004 &  119.004\\
 musae-squirrel-edges &  152.543 & 131.779 & 152.542\\
 as-skitter & 89.4696  & 71.7748  & 89.4696\\
 soc-LiveJournal1 & | & 308.551 & 320.114\\
 com-Orkut & | & 220.063 & 227.866\\
\hline
\end{tabular}
\end{table}

\section{Conclusion and Future Work}
Even though the time complexity is higher than other 2-approximation algorithms, our CBDS-P algorithm is capable of producing approximate densest subgraphs that are more accurate than those 2-approximation algorithm within a considerable time complexity. CBDS-P proposes a new approach towards calculating approximate densest subgraphs based on k-cores. 

P-Bahmani, the shared memory implementation of Bahmani et. al[] algorithm, is the first of its kind for that algorithm. Both P-Bahmani and CBDS-P are highly scalable in a shared memory, multi-core setting and it is evident by the experimental results. Further, this paper clearly expatiates the design considerations of our shared memory implementations.

Attempting to improve CBDS-P algorithm and working on shared memory implementations for other densest subgraph discovery algorithms can be identified as our future work.

%
%
%
%

\end{document}